\documentclass[
10pt,
showpacs,preprintnumbers,nofootinbib,
amsmath,amssymb,
aps,
prc,twocolumn,groupedaddress,superscriptaddress,
showkeys
]{revtex4-2}
\usepackage{epsfig, rotating}
\usepackage{multirow}
\usepackage{graphicx}
\usepackage{dcolumn}
\usepackage{bm}
\usepackage[colorlinks=true,urlcolor=blue,citecolor=blue]{hyperref}
\usepackage{color,lineno}

\begin{document}

%\title{Impact of Factorized Approximation of IMSRG(3f2) on Effective Interactions}
\title{Improving the predictive power of empirical shell-model Hamiltonians}

\author{J. A. Purcell}
\affiliation{Department of Physics and Astronomy, Michigan State University, East Lansing, Michigan 48824-1321, 
USA
and Facility for Rare Isotope Beams, Michigan State University, East Lansing, Michigan 48824-1321, 
USA}
\author{B. A. Brown}
\affiliation{Department of Physics and Astronomy, Michigan State University, East Lansing, Michigan 48824-1321, 
USA
and Facility for Rare Isotope Beams, Michigan State University, East Lansing, Michigan 48824-1321, 
USA}
\author{B. C. He}
\affiliation{ Department of Physics and Astronomy, University of Notre Dame, Notre Dame, IN 46556 
USA}
\author{S. R. Stroberg}
\affiliation{ Department of Physics and Astronomy, University of Notre Dame, Notre Dame, IN 46556 
USA}
\author{W. B. Walters}
\affiliation{Department of Chemistry and Biochemistry, University of Maryland, College Park, 
Maryland 20742, USA}

\date{\today}

\begin{abstract}
We present two developments which enhance the predictive power of empirical shell model 
Hamiltonians for cases in which calibration data is sparse. A recent improvement in the ${\it ab initio}$
derivation of effective Hamiltonians leads to a much better starting point for the 
optimization procedure. In addition, we introduce a protocol to avoid over-fitting, enabling a more reliable extrapolation beyond available data. These developments will enable more robust predictions for exotic isotopes produced at rare isotope beam facilities and in astrophysical environments.
%
%We present two effective interactions with the intention of highlighting a recent improvement in ab initio methods \cite{He24}. This work is the first paper comparing data-driven two-body matrix elements (TBME's), and concretely illustrates the importance of uncertainty in ab initio inputs to effective interactions. Given an ab initio interaction, a model space, and a set of experimental data, one can produce an improved effective interaction by fitting linear combinations of the TBME to the experimental data. Two different interactions can be constrained by data to arrive at effective interactions that produce very similar predictions; when one observes how the parameters have to change to arrive here, it becomes clear that the degree to which a Hamiltonian must be tuned to experiment depends largely on how realistic the ab initio interaction is to begin with. We show that with a recent development in ab initio methods, one can produce an effective interaction achieving a benchmark of predictive power requiring less variation in the TBME than would be required in previous methods to achieve the same predictive power. This is an indication that the new ab initio method produces an interaction that is inherently more realistic. 
\end{abstract}

\maketitle
\section{Introduction}
The nuclear shell model is a ubiquitous framework for interpreting nuclear structure data. In particular, the interacting shell model, or configuration interaction (CI) approach quantitatively reproduces and explains a vast amount of spectroscopic data. These CI calculations require the specification of an effective Hamiltonian. While ab initio methods have made great progress recently in deriving these Hamiltonians from the underlying inter-nucleon interactions, they have not yet achieved the precision obtained with phenomenological Hamiltonians adjusted to data. The gold standard for the phenomenological CI paradigm is the Universal-$sd$ (USD) family of Hamiltonians \cite{sd06, mag20}, which reproduce spectra with a root-mean-squared deviation of better than 200 keV. In this relatively small model space, the vast amount of available data is more than sufficient to constrain the parameters of the Hamiltonian. In contrast, many of the nuclei that will be studied in the coming decades at rare isotope facilities, including the majority of nuclei relevant for $r$-process nucleosynthesis, will live in larger model spaces where data are sparse. This increases the importance of maximizing predictive power with minimal data. In addition, the information content of various experimental data is often redundant, in terms of which parameters are constrained; in order to reliably extrapolate beyond the available data it is critical to avoid over-fitting.
In this paper, we (1) demonstrate that an improved ab initio calculation \cite{He24} provides a starting point which requires fewer phenomenological adjustments, and (2) utilize a training/testing partitioning scheme to avoid over-fitting.

In section II we discuss the goals and methodology of this work, including
the form of our Hamiltonian constructed from ${\it ab initio}$ methods, and the singular-value decomposition 
(SVD) method for improving the Hamiltonian by constraints to experimental data.
Section III presents the mathematical methods used for the $\chi^2$ minimization
and the SVD fitting algorithm.
Section IV is concerned with the experimental data that was used, 
including a discussion justifying the inclusion and exclusion of certain observed energy levels. 
Our results are presented in section V, and conclusions are given in section VI.
Section VII (Appendix) presents a collection of all spectra obtained with our final Hamiltonian
in comparison to experimental data.

\section{Methods}

Microscopic configuration-interaction (CI) calculations for specific regions of nuclei 
are based on a description in terms of a selected set of shell-model orbitals (the model space) 
with a  Hamiltonian operator 
\[
    H = E_0 + \sum_\alpha\epsilon_\alpha\hat{n}_\alpha + 
    \sum_{\alpha \leq \beta, \gamma \leq \delta}\sum_{J}
    V_{J}(\alpha\beta;\gamma\delta)\hat{T}_{J}(\alpha\beta;\delta\gamma)    \tag{1}
\]
that is represented by the energy of the closed core  $E_0$,
single-particle energies (SPE) $\epsilon_\alpha$, and two-body matrix elements (TBME) $V_{J}(\alpha\beta;\gamma\delta)$:
In principle, three-body matrix elements could be included, but they dramatically increase the complexity of the problem, and their main effect is to modify the SPE and TBME, so they are generally not treated explicitly.
The SPE and TBME can be obtained from a realistic NN (possibly with 3N) interaction, renormalized to the valence space in some way,
for example using many-body perturbation theory~\cite{Hjorth-Jensen1995}, shell model coupled cluster~\cite{Sun2018}, or the valence-space in-medium similarity renormalization group (VS-IMSRG)~\cite{sr19}.
We use the latter approach in this work, including the effects of 3N interactions via the normal-ordered two-body (NO2B) approximation.
This leads to values for $E_0$, $\epsilon$ and $V_J$ which are nucleus-dependent~\cite{sr19}.
%the valence-space in-medium similarity renormalization group (VS-IMSRG) approach \cite{sr19} where three-nucleon
% interactions contribute to the SPE and TBME for a given model space \cite{sr19}. The resulting values
% for $E_0$,  $\epsilon$ and  $V_{J}$
%  are nucleus-dependent \cite{sr19}.

A commonly applied approximation when optimizing empirical Hamiltonians is to use
a nucleus-independent set of SPE and TBME.
The TBME may include some smooth mass dependence, e.g. $A^{-0.3}$ \cite{sd06,mag20,fp04,jun45,jj44b}.
Within this approximation the SPE and TBME can be used as
parameters to achieve an improved description of 
measured binding energies
and excitation energies (energy data) with the goal of 
obtaining improved wavefunctions and improved predictions
for new energy data and for other observables. 
%Add citation showing that wavefunctions are indeed improved by fitting to energy data.

The singular-value decomposition (SVD) method provides a systematic approach for finding the most important linear combinations of the
$N_p$ SPE and TBME parameters which can be determined by the data \cite{Arima1968,Wildenthal1984,mag20}. One starts with wavefunctions obtained from a Hamiltonian derived from  the  best available ab-initio input.    
With these wavefunctions, the expectation value of the Hamiltonian provides a  linear combination of SPE and TBME for each of the $N_d$  energy data. The SVD amounts to a diagonalization of the $N_p \times N_p$ fit matrix solution to $\chi^2$ minimization, and provides singular values and associated eigenvectors (linear combinations of SPE and TBME). The largest singular values are associated with the most well-determined linear combinations, and smallest singular values are associated with the least well-determined combinations. The TBME associated with these least well-determined combinations will have some influence on the extrapolations to new energy data and to the calculations of other observables. One must choose a singular value cut-off $N_c$. Below the cut-off one can use TBME obtained from the best available ab-initio input. This provides a new set of SPE and TBME that can be used to obtain an improved set of wavefunctions. One iterates the SVD fits and the wavefunction calculations until convergence.

Examples of Hamiltonians obtained in this way with protons and neutrons are, 
USDA/B \cite{sd06} and USDC/I \cite{mag20} for the $\{0d_{5/2}, 0d_{3/2}, 1s_{1/2}\}$ $(sd)$ model space, 
GPFX1A \cite{fp04} for the $\{0f_{7/2}, 0f_{5/2}, 1p_{3/2}, 1p_{1/2}\}$ ($fp$) model space, 
and JUN45 \cite{jun45} and jj44b (Appendix A of \cite{jj44b}) for the $\{0f_{5/2}, 1p_{3/2}, 1p_{1/2}, 0g_{9/2}\}$ ($jj44$) model space. For all of these Hamiltonians the root-mean-square deviation (RMSD) between the calculated and experimental energy data is 150-200 keV. This is to be compared with the results of ${\it ab initio}$ type calculations over the same regions of nuclei where the RMSD is much larger (see Fig. 9 of \cite{sr19} for the $sd$ model space). 
In all of these cases, there is abundant experimental data to constrain the fitted Hamiltonian across the entire model space. However, for very exotic nuclei relevant for rare isotope beam facilities and $r$-process nucleosynthesis, the data will be sparse and strongly biased toward the most stable region of the model space. In such cases, it is possible to overfit to the available data, yielding a Hamiltonian that extrapolates poorly to more exotic nuclei.

\begin{figure}
\includegraphics[scale=0.52]{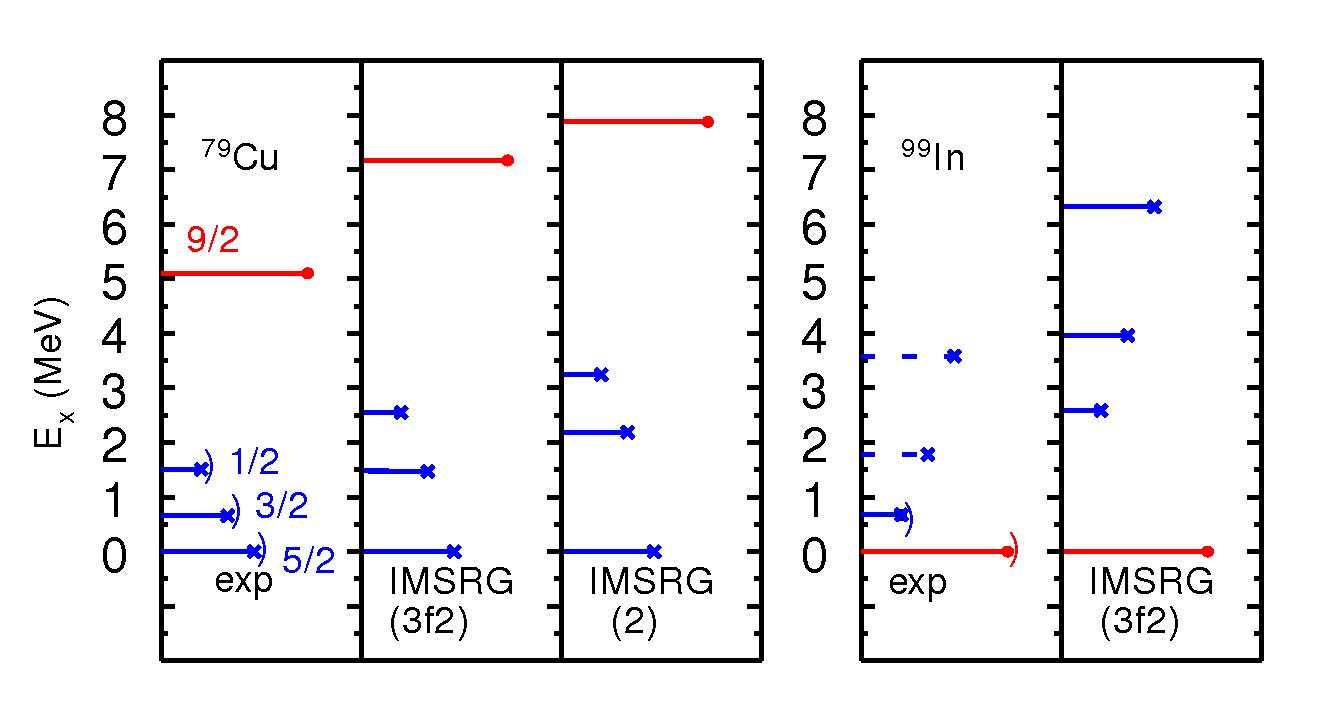}
\caption{Levels in $^{79}$Cu and $^{99}$In. 
The $J$ value is correlated with the horizontal length of the lines: 
blue for negative parity and red for positive parity. 
Experimental levels with suggested $J^{\pi}$ assignments are indicated by ")" \cite{cu79,in99}. 
The 9/2$^+$ energy shown in the experimental panel was obtained from our $p35\text{--}i3$ Hamiltonian. The dashed blue lines for $^{99}$In show excitation energies based on the observations in $^{131}$In \cite{in131}. In the $\pi j4$ model space, the states for $^{79}$Cu are interpreted as single-particle states relative to  a $^{78}$Ni closed shell, and states for $^{99}$in are interpreted as single-hole states relative to a $^{100}$Sn closed shell. The IMSRG results are based on calculations for $^{78}$Ni and $^{100}$Sn}
\label{cu79}
\end{figure}

In this paper, we demonstrate that by using an improved ab-initio starting point and by reserving some data for validation, we can improve the predictive power of the resulting Hamiltonian when extrapolating beyond the fit data.
%
%In this Letter, we consider, for the first time, uncertainties in the choice of ab-initio inputs and the choice of the cut-off. 
For our application, we consider data for low-lying states for all nuclei between $^{78}$Ni and $^{100}$Sn that can be described by protons in $\{0f_{5/2}, 1p_{3/2}, 1p_{1/2}, 0g_{9/2}\}$ (the $\pi j4$ model space), which requires 4 SPE and 65 TBME with $T=1$
associated with the valence protons.
This space has been considered previously with SVD derived Hamiltonians \cite{jun45, jw88, lisetskiy04}, as well as those obtained VS-IMSRG methods \cite{Yuan24}.
The data we employ consists of 22 ground-state binding energies and 167 excitation energies.
The region between $^{90}$Zr and $^{100}$Sn is well established territory where many wavefunctions are dominated by $\{1p_{1/2}, 0g_{9/2}\}$ configurations that requires only nine TBME that can be established directly from the energy data \cite{talmi60, cohen64, aurbach65, vervier66, ball72, gloeckner73, blomqvist85}. There is new data for nuclei close to $^{78}$Ni whose structure is dominated by the $\{0f_{5/2}, 1p_{3/2}, 1p_{1/2}\}$ subset of orbitals. Importantly, many of the
single-particle energies for these orbitals associated with low-lying states in $^{79}$Cu \cite{cu79} and  $^{99}$In \cite{in99} are now established . The results from the IMSRG methods discussed below 
are also shown in FIG \ref{cu79}. 

In section III, we review the 
experimental data for nuclei between $^{78}$Ni and $^{86}$Kr. 
This includes a discussion of intruder states that can be
attributed to orbital configurations that are not part of the $\pi j4$ model space.
These intruder states are excluded from the data set used for the SVD fits.

%The  $\pi j4$
%model space which requires 4 SPE and 65 TBME has been considered previously with SVD derived Hamiltonians \cite{jun45}, \cite{jw88}, \cite{lisetskiy04} as well as those obtained VS-IMSRG methods \cite{Yuan24}. Our results consider 22 ground binding energies and 167 
%excitation energies, an improved IMSRG-type ab-initio starting point, and a new method for determining the 
%optimal cutoff in the number varied linear combinations (VLC)
%in the SVD method.

As a starting point for our fitting procedure, we use Hamiltonians
derived with the VS-IMSRG. These are 
obtained using the EM 1.8/2.0 NN+3N interaction \cite{Heb2011} in a 
harmonic oscillator basis with frequency $\hbar \omega=12$ MeV, 
truncated to 13 major shells ($2n+l \leq$ $e_{\rm max}=12$). We normal order 
with respect to the Hartree-Fock ground state of the reference and discard the 
residual 3N interaction. We then decouple the $\pi j4$ valence space,
 using the Magnus formulation of the IMSRG. The results labeled 
IMSRG(2) are obtained with the standard approximation~\cite{sr19}, truncating all 
operators at the two-body level throughout the flow, including inside 
nested commutators. The results labeled IMSRG(3f2) include the recently 
introduced correction in which intermediate three-body operators arising 
in nested commutators are incorporated by rewriting the double 
commutator in a factorized form while maintaining the same computational 
scaling as the IMSRG(2) approximation\cite{He24}. As in ref. \cite{He24}, we include factorized terms 
with a one-body intermediate during the flow, and include terms with a 
two-body intermediate at the end of the flow. We perform the procedure for
two different references, $^{78}$Ni and $^{100}$Sn, corresponding to empty
and full valence spaces, respectively, and take the average of the two resulting
valence space Hamiltonians as our starting point for the fitting procedure. 

\begin{figure}
\centering
\includegraphics[scale=0.5]{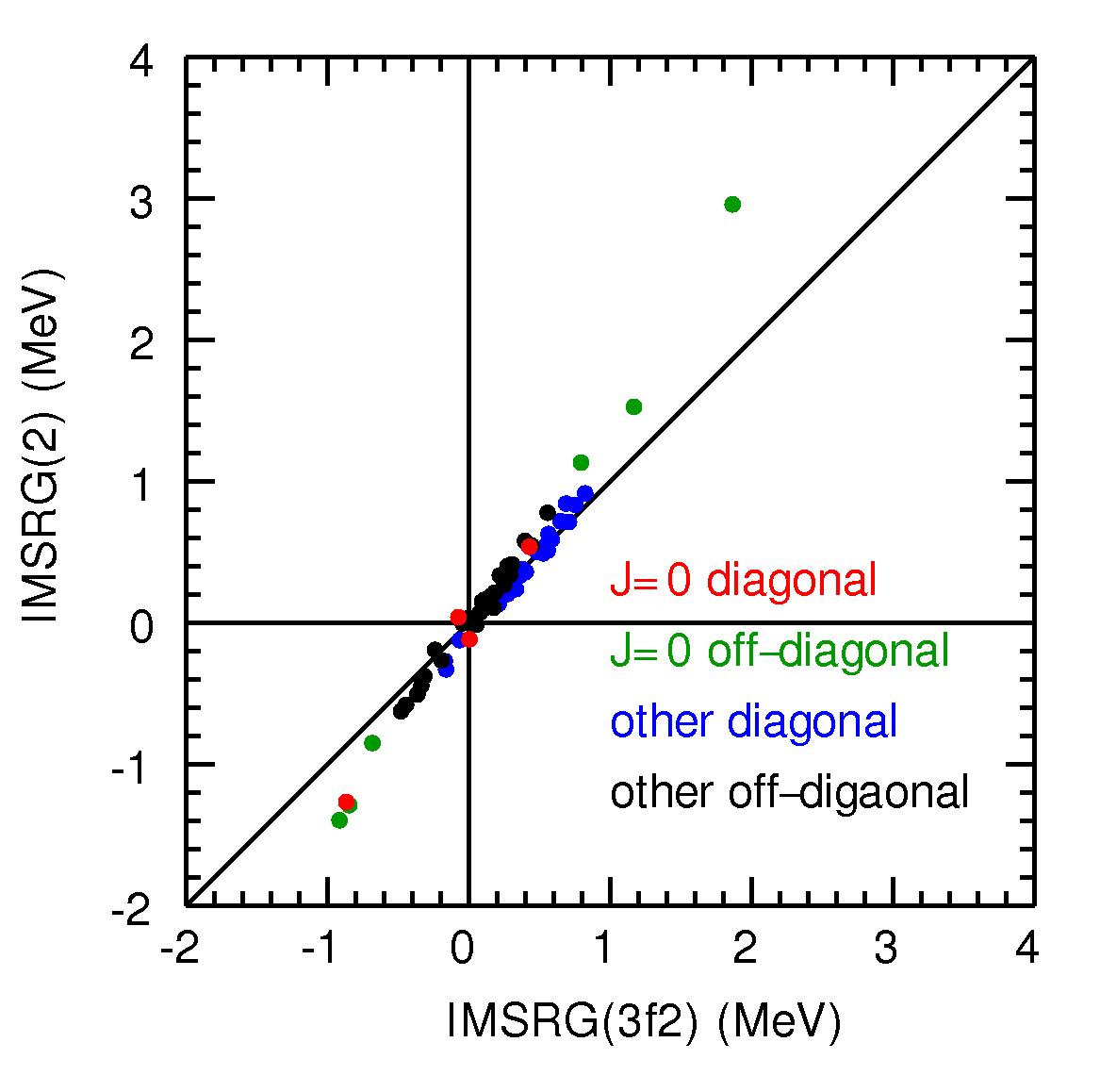}
\caption{Comparison of the TBME obtained with the two versions of
the IMSRG method - the y-axis involves the 2-body truncation and the x-axis involves the factorized approximation. The TBME are colored according to the type of overlap that they describe.
}
\label{rem78tf2}
\end{figure}

Intuitively, the more realistic the ${\it ab initio}$ Hamiltonian is, the less it needs to be modified to produce results that are in agreement with experiment.
Shown in FIG. \ref{rem78tf2} is a comparison of the TBME obtained with the IMSRG(2) and IMSRG(3f2) approximations,
indicating the magnitude of the uncertainty due to the many-body truncation.

% highlighting that the many-body methods and assumptions used can result in a distribution of ab initio interactions - in particular, FIG. \ref{rem78tf2} gives a snapshot of the uncertainty in ab initio methods.
 These uncertainties propagate into our effective Hamiltonians, and ultimately the resulting wavefunctions and calculations made with them. 
The main difference is that the $J=0$, $T=1$ TBME are about 30 percent
weaker for IMSRG(3f2) compared to those from IMSRG(2).

%The propagation of these ab initio uncertainties into effective interactions such as these is a computationally intensive current research problem.

\section{Math Methods}

\subsection*{\texorpdfstring{Initial Procedure: $\chi^2$ Minimization}{L2}}
Our starting Hamiltonian has a set of parameters $\Vec{p}^s = (p_1^s,\cdots, p_{N_p}^s)$. This Hamiltonian defines a starting set of eigenvectors $|\phi_k^s\rangle$ that can be used to calculate the operator overlaps $\beta_i^k$ so that each associated eigenvalue $\lambda_k$ can be calculated in equation $(1)$. Then we seek to minimize the $\ell^2$-norm of the residual ($\Vec{\chi}$) between the $N_d$ measured experimental energies and calculated energy eigenvalues:
\[
    \underset{\Vec{p}\in \mathbb{R}^{N_p}}{\textrm{min}}\chi^2 = 
    \underset{\Vec{p}\in \mathbb{R}^{N_p}}{\textrm{min}}
    \sum_{k=1}^{N_d}
    \left(\frac{E_k^{exp}-\lambda_k(\Vec{p})}{\sigma_k}\right)^2
\]
where $\sigma_k^2 = \left(\sigma_k^{exp}\right)^2 + \left(\sigma_k^{th}\right)^2$. We can reorganize the fitting by expanding $\lambda_k$ and defining the expected energy contribution from $H^1(\Vec{p})$ to be $\epsilon_k^{exp} = E_k^{exp} - E_k^0$. Now our $\chi^2$ minimization looks like:
\[
    \underset{\Vec{p}\in \mathbb{R}^{N_p}}{\textrm{min}}\chi^2 = 
    \underset{\Vec{p}\in \mathbb{R}^{N_p}}{\textrm{min}}
    \sum_{k=1}^{N_d}
    \left(\frac{\epsilon_k^{exp}-\epsilon_k(\Vec{p})}{\sigma_k}\right)^2
\]
We simplify the notation further first by re-expressing our data components $z_k^{exp} = \epsilon_k^{exp}/\sigma_k$ and then considering the model components:
\[
    z_k(\Vec{p}) = \frac{\epsilon_k(\Vec{p})}{\sigma_k} = \sum_{i=1}^{N_p}p_i\frac{\beta_i^k}{\sigma_k}
\]
We can arrange the components of the experimental data into a data vector $\Vec{z}^{exp} = [z_1^{exp},\cdots,z_{N_d}^{exp}]^T$, and similarly for the model components we can represent a model vector through matrix-vector multiplication:
\[
    \left[\begin{array}{c}
        z_1(\Vec{p}) \\
        \vdots \\
        z_{N_d}(\Vec{p})
    \end{array}\right] = 
    \left[\begin{array}{ccc}
        \beta_1^1/\sigma_1 & \cdots & \beta_{N_p}^1/\sigma_1 \\
        \vdots & \ddots & \vdots \\
        \beta_1^{N_d}/\sigma_{N_d} & \cdots & \beta_{N_p}^{N_d}/\sigma_{N_d} \\
    \end{array}\right]
    \left[\begin{array}{c}
        p_1 \\ \vdots \\ p_{N_p}
    \end{array}\right]
\]
If we recognize that $\frac{\beta_j^k}{\sigma_k} = \frac{\partial z_k(\Vec{p})}{\partial p_j}$ then the matrix above must by definition be the transposed \underline{Jacobian} $(J^T)$ of the vector-valued function $\Vec{z}(\Vec{p}) = [z_1(\Vec{p}),\cdots,z_{N_d}(\Vec{p})]^T$. This allows us to re-express our $\chi^2$ minimization as:
\[
    \underset{\Vec{p}\in \mathbb{R}^{N_p}}{\textrm{min}}\chi^2 = 
    \underset{\Vec{p}\in \mathbb{R}^{N_p}}{\textrm{min}}\Vert\Vec{z}^{exp} - J^T\Vec{p}\Vert^2
\]
Minimizing this with respect to each parameter $p_j$ gives $N_p$ equations of the form:
\begin{align*}
    \frac{\partial \chi^2}{\partial p_j} &= 
    2\sum_{k=1}^{N_d}\left(\frac{\epsilon_k^{exp}-\epsilon_k(\Vec{p})}{\sigma_k}\right)
    \left(-\frac{1}{\sigma_k}\frac{\partial\epsilon_k(\Vec{p})}{\partial p_j}\right) \\ &=
    -2\sum_{k=1}^{N_d}\left(\frac{\epsilon_k^{exp}\beta_j^k}{\sigma_k^2} - \sum_{i=1}^{N_p}p_i\frac{\beta_i^k\beta_j^k}{\sigma_k^2}\right) = 0
\end{align*}
which leaves us with the condition that:
\[
    \sum_{k=1}^{N_d}\frac{\epsilon_k^{exp}\beta_j^k}{\sigma_k^2} = 
    \sum_{k=1}^{N_d}\sum_{i=1}^{N_p}p_i\frac{\beta_i^k\beta_j^k}{\sigma_k^2}
\]
Notice that 
If we now define the quantities:
\[
    e_j = \sum_{k=1}^{N_d}\frac{E_{exp}^k\beta_j^k}{\left(\sigma^k\right)^2} 
    \quad \textrm{and} \quad
    G = \left[\begin{array}{ccc}
         \gamma_{11} & \cdots & \gamma_{1p}  \\
         \vdots & \ddots & \vdots \\
         \gamma_{p1} & \cdots & \gamma_{pp}
    \end{array}\right] 
\]
where
\[
    \gamma_{ij} =
    \sum_{k=1}^{N_d}\frac{\beta_i^k\beta_j^k}{\left(\sigma^k\right)^2} = \sum_{k=1}^{N_d}\frac{\beta_j^k\beta_i^k}{\left(\sigma^k\right)^2} = \gamma_{ji} \in \mathbb{R}  \tag{$G = G^T$}
\]

we can rewrite the $N_p$ equations as a single vector equation:
\[
    \Vec{e} - G\Vec{x} = 0 \quad \longrightarrow \quad
    \Vec{x} = G^{-1}\Vec{e}
    \tag{2}
\]
Since the $G$-matrix is real and symmetric, it is diagonalizable. The step of solving for $\Vec{x}$ shown in equation $(2)$ is only possible if G is invertible, meaning that none of the eigenvalues of the $G$-matrix are 0. This process can be used to find a new $\Vec{x}$ that can be used to calculate new eigenvalues for the new eigenstates $|\phi_k'\rangle $, repeating until convergence.$\hfill \break$
$\mathbf{Note:}$ $G^{-1}$ is referred to as the error matrix because its diagonal entries are the square of parameter errors and the off-diagonals are related to correlations between parameters.

\subsection*{SVD Proceedure}
The Hamiltonian parameters are often highly correlated, and the fit can be re-expressed in terms of an orthonormal basis of uncorrelated \underline{SVD parameters}. Since $G$ is real and symmetric, its SVD is identical to an eigendecomposition - the SVD for $G$ and $G^{-1}$ are:
\[
    G = ADA^T \quad \textrm{and} \quad
    G^{-1} = AD^{-1}A^T
\]
where:
\begin{itemize}
    \item[-] $D \in \mathbb{R}^{p \times p}$ is a matrix of diagonal positive elements $D_{ii} > 0$.
    \item[-] $A \in \mathbb{R}^{p\times p}$ is a rotation matrix whose columns form an orthonormal basis of our SVD parameter space.
    \item[-] $D^{-1}$ is a diagonal matrix whose elements are inverses of the elements of $D$ - i.e. $\left[D^{-1}\right]_{ii} = d_i = \frac{1}{D_{ii}}$
\end{itemize}
With these substitutions in the result of our minimization we have:
\[
    \Vec{x} = AD^{-1}A^T\Vec{e} \quad \longrightarrow \quad
    A^T\Vec{x} = D^{-1}A^T\Vec{e}
\]
Now we express the rotations of $\Vec{x}$ and $\Vec{e}$  
\[
    \Vec{y} = A^T\Vec{x}
    \quad \textrm{and} \quad
    \Vec{c} = A^T\Vec{e}
\]
resulting in
\[
    \Vec{y} = D^{-1}\Vec{c} \quad \longrightarrow \quad 
    y_i = d_ic_i \tag{3}
\]
Here the uncorrelated SVD parameters $y_i$ are expressed as a linear combination of the Hamiltonian parameters $x_i$ with associated errors $d_i$ - explicitly:
\[
    \Vec{y} = A^T\Vec{x} = \sum_{l=1}^px_l\Vec{a}_l^T 
    \quad \longrightarrow \quad
    y_i = \sum_{l=1}^px_l\left[A^T\right]_{il} 
    \tag{4}
\]
where $\Vec{a}_l^T$ is the $l$'th column of $A^T$. When $d_i$ is large the SVD parameters $y_i$ experience a large change from a correspondingly small change in the data $c_i$, meaning that the corresponding linear combination of Hamiltonian parameters $y_i$ is poorly determined by the given data set. We can establish a cutoff criterion on what is poorly determined linear combination of Hamiltonian parameters $y_i$ based on the magnitude of the corresponding $d_i$

\subsection*{Fitting Algorithm}
\begin{enumerate}
    \item Starting from the best available Hamiltonian parameters $\Vec{x}^s$ we construct and diagonalize the $G$-matrix to obtain $D_{ii}$ eigenvalues and the orthonormal basis for our parameter space. 
    \item Mutually independent \underline{SVD parameters} $y_i$ are determined in the fit from equation $(4)$. Explicitly:
    \[
        y_i = d_ic_i = d_i\sum_{l=1}^pe_l\left[A^T\right]_{il}
    \]
    Simultaneously, linear combinations of ${\it ab initio}$ \underline{Hamiltonian parameters} are determined from equation $(4)$. Explicitly:
    \[
        \Vec{y}^* = A^T\Vec{x}^* \quad \longrightarrow \quad
        y_i^* = \sum_{l=1}^px_l^*\left[A^T\right]_{il}
    \]
    \item One defines a cutoff criterion $\delta$, and updated linear combinations $\Vec{y}^a$ are defined by only adopting well-determined values $y_i$ with respect to this cutoff, and leaving ${\it ab initio}$ values $y_i^*$ for the rest:
    \[
        y_i^a = \begin{cases}
            y_i \quad &(d_i \leq \delta) \\
           y_i^* \quad &(d_i > \delta)
        \end{cases}
    \]
    The number of well-determined linear combinations is $N_d$
    \item With the updated set of model parameters $\Vec{y}^a$ one recovers the Hamiltonian parameters by inverting the rotation:
    \[
        \Vec{x}^a = (A^T)^{-1}\Vec{y}^a
    \]
    \item This set of Hamiltonian parameters takes the place of $\Vec{x}^s$ as input to the first step of this algorithm, and is used to obtain the next set of parameters $\Vec{x}^b$. This process is repeated until convergence.
\end{enumerate}

\section{Experimental data for N = 50 isotones}

For the data used as input for
the SVD fits we chose levels which have reliable
excitation energies and $J^{\pi}$ values as determined
by various types of experiments. In addition,
there are levels not included in the SVD fits;
those with uncertain $J^{\pi}$ values and
those that can be considered as intruder states
into the $\pi j4$ model space.
We use experimental data from 
the Evaluated Nuclear Structure Data Files (ENSDF) \cite{ensdf}.
The purpose of this section is to review the relatively
new data for the nuclei  $^{78}$Ni up to $^{86}$Kr 
with regard to these criteria for inclusion. 
The subsequently chosen data are 
given in \cite{supp}.

The structure of $^{78}$Ni has been studied via knock-out reactions
and shows a group of excited states between 2.6 and 4.0 MeV
\cite{ni78}.
Levels with spins and parities of
2$^{ + }$ and 4$^{ + }$ are suggested at 2.60 and 3.18 MeV, respectively.
The relatively high excitation of these states indicates
that the wavefunction of $^{78}$Ni is
dominates by closed-shell configuration
of $  Z=28  $ for protons and $  N=50  $ for neutrons.
For protons, a $  0f_{7/2}  $ proton can be moved
across the $  Z=28  $ proton gap into the
$  0f_{5/2}  $ and $  1p_{3/2}  $ orbitals leading to a
multiplet of states with $  J^{\pi }  $ ranging from 1$^{ + }$ to 6$^{ + }$.
For neutrons, a $  0g_{9/2}  $ neutron can be moved
across the $  N = 50  $ shell gap
into the $  0d_{5/2}  $ orbital, leading to a multiplet of states
states from
2$^{ + }$ to 7$^{ + }$.  Also for neutrons, a $  1p_{1/2}  $ neutron can
be moved across $  N=50  $ into the $  0d_{5/2}  $ orbital, leading to
states with 2$^{-}$ and 3$^{-}$.

The level structure of $^{79}$Cu has also been studied in
knock-out reactions \cite{cu79}.
Two low-lying levels are proposed at 0.656 and 1.511 MeV
with the gamma-decay sequence of
1.511 MeV to 0.656 MeV to the ground state.
These experimental energies are compared to
shell-model predictions in FIG. \ref{cu79}.  All of these predictions
give an ordering of 1/2$^{-}$, 3/2$^{-}$ and 5/2$^{-}$ corresponding
to single-particle configurations of $  1p_{1/2}  $, $  1p_{3/2}  $
and $  0f_{5/2}  $, respectively.

The $^{79}$Cu ground-state spin-parity is not measured.
The  lighter Cu isotopes are measured to have $^{71}$Cu ($  J=3/2  $),
$^{73}$Cu ($  J=3/2  $), $^{75}$Cu ($  J=5/2  $) and $^{77}$Cu ($  J=5/2  $)
\cite{ensdf}.
Shell-model calculations give negative-parity ground states.
The energy differences between the lowest levels with $  J=3/2^{-}  $
and $  J=5/2^{-}  $ obtained with the JUN45 Hamiltonian \cite{jun45}
in the $  jj44  $ model space are shown in Fig. (6). They are
compared with experiment for $^{69,71,73,75}$Cu
using the energies of the states
that are suggested to have $  J  $ = 3/2 and 5/2 \cite{ensdf}.
For $^{79}$Cu we give the experimental results assumming a ground state and
first excited states with $  J^{\pi }  $ of 5/2$^{-}$ and 3/2$^{-}$, respectively,
(solid black line) and 3/2$^{-}$ and 5/2$^{-}$, respectively.
(dashed black line).
The systematics of these energy differences compared to theory
are consistent with the ground state of $^{79}$Cu having
$  J^{\pi }  $=5/2$^{-}$, with the first excited state at 0.656 MeV
having $  J^{\pi }  $=3/2$^{-}$. When compared to the shell-model
calculations in Fig. (1), the second-excited
state at 1.511 MeV is expected to have  $  J^{\pi }  $=1/2$^{-}$.
The sequence of
gamma decays observed in \cite{cu79} is consistent with
the (1/2$^{-}$, 3/2$^{-}$, 5/2$^{-}$) sequence.
In the single-particle model this gives a $  1p_{1/2}-1p_{3/2}  $
spin-orbit splitting of 0.855 MeV.
This smaller the value of 1.66 MeV shown by the MCSM calculations
in FIG. 2 of \cite{cu79}. The experimental 1/2$^{-}$ - 3/2$^{-}$
splitting in in $^{131}$In is suggested to be 
%\sout{1.35 MeV} 
988 keV \cite{tap16}.

Other excited states in $^{79}$Cu are observed starting at 2.9 MeV.
It is natural to understand these as intruder coming from the knockout
of an $  0f_{7/2}  $ proton in $^{80}$Zn leading to $  2p-1h  $ states
in $^{79}$Cu.
Three transitions at 2.94, 3.88, and 4.30 MeV  of about equal intensity
that decay directly to the ground state and could represent
fragments of the $  0f_{7/2}  $ hole strength.

The first excited 2$^{ + }$ level in $^{80}$Zn was identified
in a Coulomb excitation study at 1.492 MeV \cite{zn80w}.
The Coulex experiment obtains B(E2,$  \uparrow  $)=20.1(16) e$^{2} \cdot$fm$^{4}$.
Further level structure has been provided from knock-out reactions
where a 4$^{ + }_{1}$ level is placed at 1.93MeV \cite{zn80s}.
In \cite{zn80s} three other levels are identified at
2.627, 2.820 and 3.174 MeV. The latter two decay only to the 4$^{ + }_{1}$ level.
The experimental energies are compared to those from the JUN45 \cite{jun45}
and MCSM \cite{mcsm} Hamiltonians in FIG. 4 of \cite{zn80s}.
With only two protons in the $  0f_{5/2}  $ and $  1p_{3/2}  $
orbitals beyond $^{78}$Ni, the maximum positive-parity spin is 4$^{ + }$.
Lifetimes and B(E2) values have been determined for both the $2^+$ and $4^+$ levels \cite{zn80cort}.

Level structure for $^{81}$Ga has come from extensive study of
the beta decay of the 5/2$^{+}$ ground state of  $^{81}$Zn,
as well as by both multi-nucleon transfer (MNT) and knock-out reactions.
A recent decay scheme was obtained using laser-ionized $^{81}$Zn
and has provided the most detail \cite{ga81p}.
The ground-state 5/2$^{-}$ assignment is supported by laser hyperfine methods \cite{ch10}.
Two excited 3/2$^{-}$ levels are identified, along with one
clear 9/2$^{-}$ level and one clear 11/2$^{-}$ level \cite{ga81d}.
11/2$^{-}$ is the highest spin that can be obtained by three protons
in the $0f_{5/2}$ and $1p_{3/2}$ orbitals.
Levels proposed at higher spins  would have to involve either the
$  0g_{9/2}  $ proton orbital, or intruder states coming from the
neutron excitations across the $  N=50  $ shell gap.
A possible higher-spin level has been reported by both \cite{ga81d} and \cite{ga81guil} at 2.766 MeV that decays to the 11/2$^{-}$ level. Ref. \cite{ga81d} also reports on the gamma decay from a level at
3.093 MeV to the 2.766 MeV level, that could have even higher spin. Transitions at 770 keV and 990 keV were observed in (p,2p) reactions that feed into the $11/2^-$ level at 1.952 MeV \cite{olivPhD}.  

The basic level structure for $^{82}$Ge has been provided by beta-decay of the 2$^{-}$ ground state of $^{82}$Ga \cite{ge82alsh, ge82vern}, and beta-delayed neutron decay of the 5/2$^{-}$ ground-state of $^{83}$Ga. Also by Coulomb excitation, MNT studies \cite{ge82rzac, ge82hwan, ge82carp, ge82sahi}, and fission-product gamma-ray studies \cite{ge82wilm}. Triple coincidences between gamma rays (in MeV) at 1.348 (2$^{ + }$), 0.938 (4$^{ + }$), and 0.940 (6$^{ + }$) establish the yrast sequence. As $6^+$ is the highest possible spin for four protons in the $0f_{5/2}$ and $1p_{3/2}$ levels, this level would have to arise from some cross-shell excitation, or involvement of the $0g_{9/2}$ proton orbital. A  proposed $7^+$ level at 3.948 MeV has been reported \cite{thisse23} that decays to the $6^+$ level at 3.228 MeV. These $5^+,6^+,7^+$ states are understood as coming from
the excitation of a neutron in the $0g_{9/2}$ orbital into the $1d_{5/2}$ orbital across the $N=50$ shell
gap \cite{thisse23,sie12}. These high-spin particle-hole configurations are also observed in 
 $^{84}$Se, $^{86}$Kr and $^{88}$Sr  \cite{thisse23} at increasingly
 higher excitation energy.

The 2$_{2}^{ + }$ MeV level is placed at 2.216 MeV by strong population in all of the decay studies as well as the presence of a
2.216 MeV ground-state transition. The 0$_{2}^{ + }$ level at 2.333 MeV  is similarly observed in decay studies with no ground-state transition. Two levels at 2.702 and 2.714 MeV are observed in all of the decay studies, but not observed in the MNT and fission-gamma studies are possible 3$^{ + }$ and 1$^{ + }$ levels, respectively. Two levels at 2.883 and 2.933 MeV are observed in decay, MNT, and fission studies that decay only to the two lower-energy 4$_{1}^{ + }$ and 4$_{2}^{ + }$ levels and are candidates for 4$^{ + }$ and 5$^{ + }$ assignments.  The B(E2) value for the 1.348 MeV level has been measured by \cite{ge82gade}.

The level structure of $^{83}$As with 33 protons has the most
complex structure among the odd-proton $  N=50  $ isotones.
It has 5 valence protons and is exactly half way from $  Z = 28  $ to
$  Z = 38  $.  The maximum spin available by breaking both pair of protons
that occupy the $  0f_{5/2}  $ and $  1p_{3/2}  $  orbitals would be 13/2$^{-}$.
Extensive data exist for the structure of $^{83}$As from decay and MNT reaction studies.
With 5 protons, the Fermi level has moved up in energy to the point
that a 9/2$^{ + }$ state associated with the $  0g_{9/2}  $ proton
state should be observed.
However, firm identification has remained elusive.
A candidate is present at 2.777 MeV that has been assigned (9/2$^{ + }$).
\cite{as83bacz}.
However, in other studies, no level was assigned as 9/2$^{ + }$
\cite{as83rezy, se84drou, as83porq}.
The gamma decay data clearly show low-energy levels that
have should have proton configurations,
then a gap around 3 MeV where the neutron particle-hole states should be
present \cite{as83winger}.
As in $^{81}$Ga, three excited levels in $^{83}$As at
0.307, 1.544, and 1.867 MeV are clearly identified as
3/2$^{-}$, 9/2$^{-}$, and 11/2$^{-}$, respectively.

Extensive data are also present for the structure of $^{84}$Se.
Not only from beta decay and multi-nucleon transfer (MNT) studies,
but also from the $^{82}$Se(t,p)$^{84}$Se 2-neutron transfer reaction \cite{se84mullin},
and where cross-shell neutron states are possible,
the level density rises rapidly. Neutron excitations across the shell gap give rise to intruder states, and one expects similar intruder states in even-even nuclei across the region of interest. 
Levels populated in beta decay were reported by
Hoff {\it et al.} \cite{se84hoff}.
High-spin structures have been reported by several groups
\cite{ge82carp, se84drou}.

Three prominent  higher-spin levels have been observed at 3.372 [6$^{ + }$],
3.639 [5$^{ + }$], and 3.704 [6$^{ + }$] MeV with the spin and parity
assignments derived from respective Coulex data \cite{se84litz}, \cite{se84mullin}, \cite{se84knight}. A proposed $7^+$ level has been identified in several papers at 4.407 MeV.  Unlike $^{82}$Ge, it is possible to obtain spin and parity of $7^+$ by aligning all six protons in the $0f_{5/2}$, $1p_{3/2}$, and $1p_{1/2}$ orbitals \cite{thisse23}.
Two 0$^{ + }$ levels below 3 MeV are identified in both reports at 2.247
and 2.655 keV.  Gamma-ray transitions from both of these
levels are seen by cross-correlations in the Carpenter data set \cite{ge82carp}.
Three other 0$+$ levels are proposed at 1.967, 2.716, and 2.740 MeV
for which no gamma transitions are observed.

The level structure of $^{85}$Br is quite important for these studies as the
data for the spins and parities are on a far firmer basis.
Polarized proton pickup in the $^{86}$Kr(d,$^{3}$He)$^{85}$Br reaction
provide definitive data for the locations of the
3/2$^{-}$, 5/2$^{-}$ and 1/2$^{-}$ states at 0, 0.345, and 1.191 MeV, respectively.
These are associated with $  1p_{3/2}  $, $  0f_{5/2}  $, and
$  1p_{1/2_{}}  $ single proton hole states.
Angular distribution data for the high-spin levels also provide a
definite location of the 9/2$^{ + }$ level at 1.859 MeV
that is associated with proton excitation into the $  0g_{9/2}  $ orbital. The yrast $9/2^-$ and $11/2^-$ levels observed in the lower-Z isotones are clearly identified at 1.572 and 2.165 MeV, respectively.

Excellent data are available for the levels of $^{86}$Kr.
The new beta-decay data from the 1$^{-}$ $^{86}$Br ground state
leads directly to a certain 2$^{-}$ assignment for the level at 4.316 MeV \cite{kr86urban}.
There is also good agreement for the high-spin levels from studies
by two different groups \cite{kr86prev, kr86winter}.
There are Coulex data for the 2$^{ + }$ and 3$^{-}$ levels and a surprising long
3-ns half-life for the 4$_{1}^{ + }$ level.

The data chosen for the SVD fit are given  in \cite{supp}.
The majority of these data
have experimental uncertainties of a few keV. 
The exceptions are: 100 keV for the $^{79}$Cu ground state
\cite{wang2021},
77 keV for the $^{99}$In ground state \cite{in99mou}, 
37 keV for the $^{99}$In first excited ($1/2^-$) \cite{in99},
and 240 keV for the $^{100}$Sn ground state \cite{sn100hin,wang2021}.
In order to constrain the SPE at the beginning and the end
of the $\pi j4$ model space, smaller uncertainties of 
50 keV (without the theoretical error of 150 keV) were 
used for the $3/2^-$ and $1/2^-$ 
excited states observed for $^{79}$Cu and for the $3/2^-$ and $5/2^-$
excited states extrapolated for $^{99}$In.  

\begin{figure*}[h!tbp]
\centering
    \centering
    \includegraphics[scale=0.32]{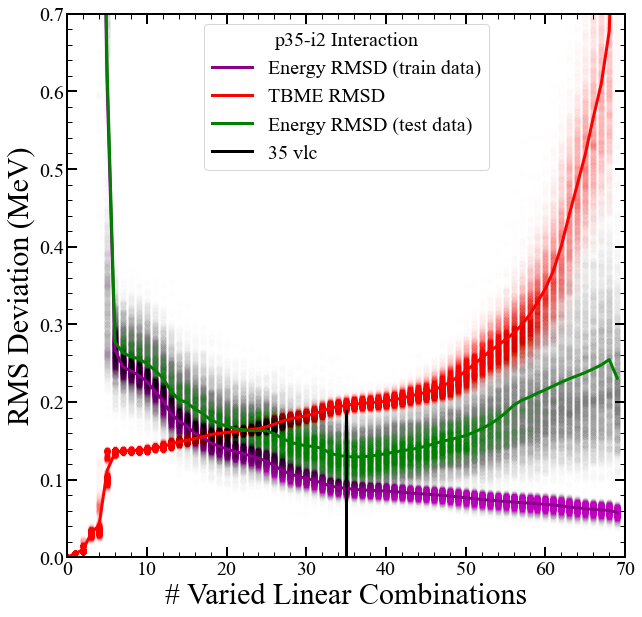}
    \includegraphics[scale=0.32]{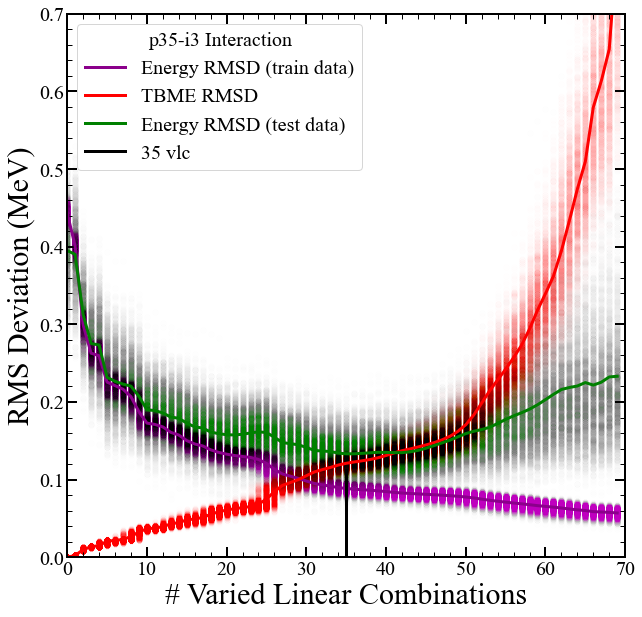}

\caption{
Results of the SVD fits as a function of the number of varied linear combinations (VLC)
of parameters. The left-hand side shows results used to obtain the p35-i2 Hamiltonian,
and the right-hand side shows results used to obtain the p35-i3 Hamiltonian.
The purple points show the energy-RMSD between theory and experiment
as a function of the number of VLC, and the red points show the TBME-RMSD between the IMSRG and fitted Hamiltonians.
The lines are the results for the pn-i2 and pn-i3 Hamiltonians.
We also show points for the 
2000 batches of randomly sampled experimental training data 
discussed in the text, purple points for the training energy data set,
and green points for the predicted data set.
}
\label{batches}
\end{figure*}

\section{Results}

The theoretical uncertainty $\sigma_{th}$ is taken to be 150 keV.
Most experimental uncertainties are on the order of a few keV,
with a larger uncertainties of a few hundred keV in the some binding energies.
The value of the $\sigma_{th}$ is chosen to give a $\chi^2$ of about unity
for the final fit. This effectively makes the weights of all energy
data about equal.
Changing $\sigma_{th}$  in the range of 100-200 keV has very minimal
effect on the results.

FIG. \ref{batches} shows the RMSD for the two different IMSRG starting points as a function of the number of varied linear combinations (VLC) of parameters. The
 resulting Hamiltonians will be labeled by
 $pn\text{--}ia$ where $n$ indicates the number of VLC,
 and $a$ indicates the ab-initio starting Hamiltonian, 
 $a=2$ for IMSRG(2) and $a=3$ for IMSRG(3f2). 
 The purple line shows the energy-RMSD between the calculated and
 experimental energy data. 

\begin{figure}
\includegraphics[scale=0.32]{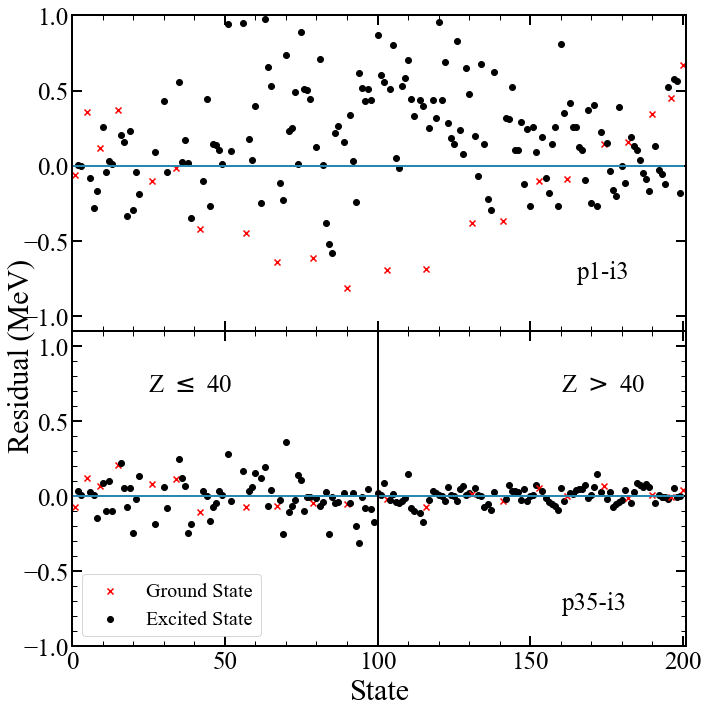}
\caption{Residual between states calculated from effective Hamiltonian with both 
1 VLC (p1-i3) and 35 VLC (p35-i3), and corresponding experimental states used to generate the 
effective Hamiltonian. Both effective Hamiltonians are obtained from the IMSRG(3f2) starting point.
We note that the  p35-i3 energy differences are larger for $Z \leq 40$ (120 keV) compared 
those above that (50 keV).
The main reason is that the higher-mass data is dominated by only  nine TBME 
associated with the $\{1p_{1/2}, 0g_{9/2}\}$ orbitals, whereas the lower mass data
is dominated by 30 TBME associated with the $\{0f_{5/2},1p_{3/2},1p_{1/2}\}$ orbitals.}
%The supplementary material includes a listing of the experimental and p35-i3 fitted energy data
%as well as a listing of the p35-i3 SPE and TBME.}
\label{residuals}
\end{figure}

Both starting points approach a similar level of energy-RMSD as the number of VLC increase. The major difference between 
 the approach to this threshold is shown by the red points, which shows 
 the RMSD between the fitted and ab-initio TBME.
 For IMSRG(2) the TBME-RMSD
 goes above the 200 keV level with 6 VLC,
 while for the results for IMSRG(3f2) do not reach this level until around 55 VLC. It can also be seen that the energy-RMSD for the fitted
 data set falls substantially after only one VLC for IMSRG(3f2), while 
 IMSRG(2) requires nearly 15 VLC to reach the same level of accuracy. These results highlight the improvement made for the TBME by the IMSRG(3f2) method.
  The deviations for each state in the fitted energy data obtained with the p1-i3 and p35-i3 Hamiltonians 
are shown in FIG. \ref{residuals}. The IMSRG(3f2) Hamiltonian should provide
better input for the undetermined SVD linear combinations compared to IMSRG(2).
The SPE and TBME for the p35-i3 Hamiltonian, and the numerical 
results of the p35-i3 SVD fit in comparison to experimental data are given \cite{supp}.

%Within the VS-IMSRG approach the $E_0$,  SPE and TBME of Eq. 1
%are all nucleus dependent.
%Thus, the 15 percent change between the overall monopole interaction
%from  the TBME obtained from IMSRG(3f2)  for 78Ni and our  p35-i3  fit
%for binding energies up to 100Sn,
%could be attributed to a small nucleus-dependent
%change in the $E_0$ and/or the SPE.  
%More work needs to be done to understand the relative contributions
%of these terms.

%\begin{figure*}
%\centering
%    \centering
%    \subfloat{\includegraphics[scale=0.32]{First Paper/r35RMSE.png}}
%    \subfloat{\includegraphics[scale=0.32]{First Paper/t35RMSE.png}}

%\caption{RMSE for both interactions. The red curve gives the RMSE between the TBME before and after fitting a given number of linear combinations of Hamiltonian parameters, while the purple gives the RMSE between theory and experiment before and after fitting.
%}
%\label{ (2)] }
%\end{figure*}

An essential purpose of these effective Hamiltonians is to achieve some level of predictive power; this is highlighted in FIG. \ref{batches}. To obtain the scatter of the points around the central fit values
as well as the green points in this figure, all of the known experimental data for this model space was randomly partitioned into training (80\%) and testing (20\%) batches. The training batch was used to vary the parameters of our Hamiltonian, and the testing set was used in the calculation of RMSD - in this way our fitted Hamiltonian is predicting the results of data that it had not seen in the fitting procedure. The sampling process was repeated 2000 times to generate a distribution of calculations for each VLC, and for each set of
Hamiltonian parameters. The green curve highlights the predictive capacity of each Hamiltonian as the number of VLC's is increased. It can be seen that each Hamiltonian approaches an energy-RMSD minimum 
(a maximum for the predictive power) at about 35 VLC. 
In fact the RMSD is reasonably small over the range of 15-35 VLC. 
This is significant because it indicates that we can achieve some benchmark level of predictive power with fewer modifications to our starting Hamiltonian with the factorization method given by \cite{He24}. This becomes critically significant when one considers effective Hamiltonians for larger model spaces where producing these effective Hamiltonians and using them to make predictions becomes computationally intensive, or where data is sparse or redundant in terms of constrained parameters.

%For p35-i2, not only is the parameter RMSD much larger, but the minimum in testing RMSD is more well-defined at 35 VLC. 
%For the p35-i3 interaction we can see that the green curve is centered much more gradually about 35 VLC, with the smaller numbers of VLC approaching similar results in terms of predictive power.
%The p35-i2 interaction has deviated much more from it's start than the p35-i3 interaction has - the TBME residual before and after the fit for the IMSRG(3f2) interaction is shown in FIG. \ref{initialFinalt}.

%The final result of our fit is contained within the distribution of points on FIG. \ref{batches}, it is the curve one would generate when every single well-determined nuclear state is included in the fit as testing data.

%FIG. \ref{r35ct35b} illustrates that while these two interactions arrive at similar predictions, the parameters of these final interactions are very different. 

\begin{figure}
\includegraphics[scale=0.52]{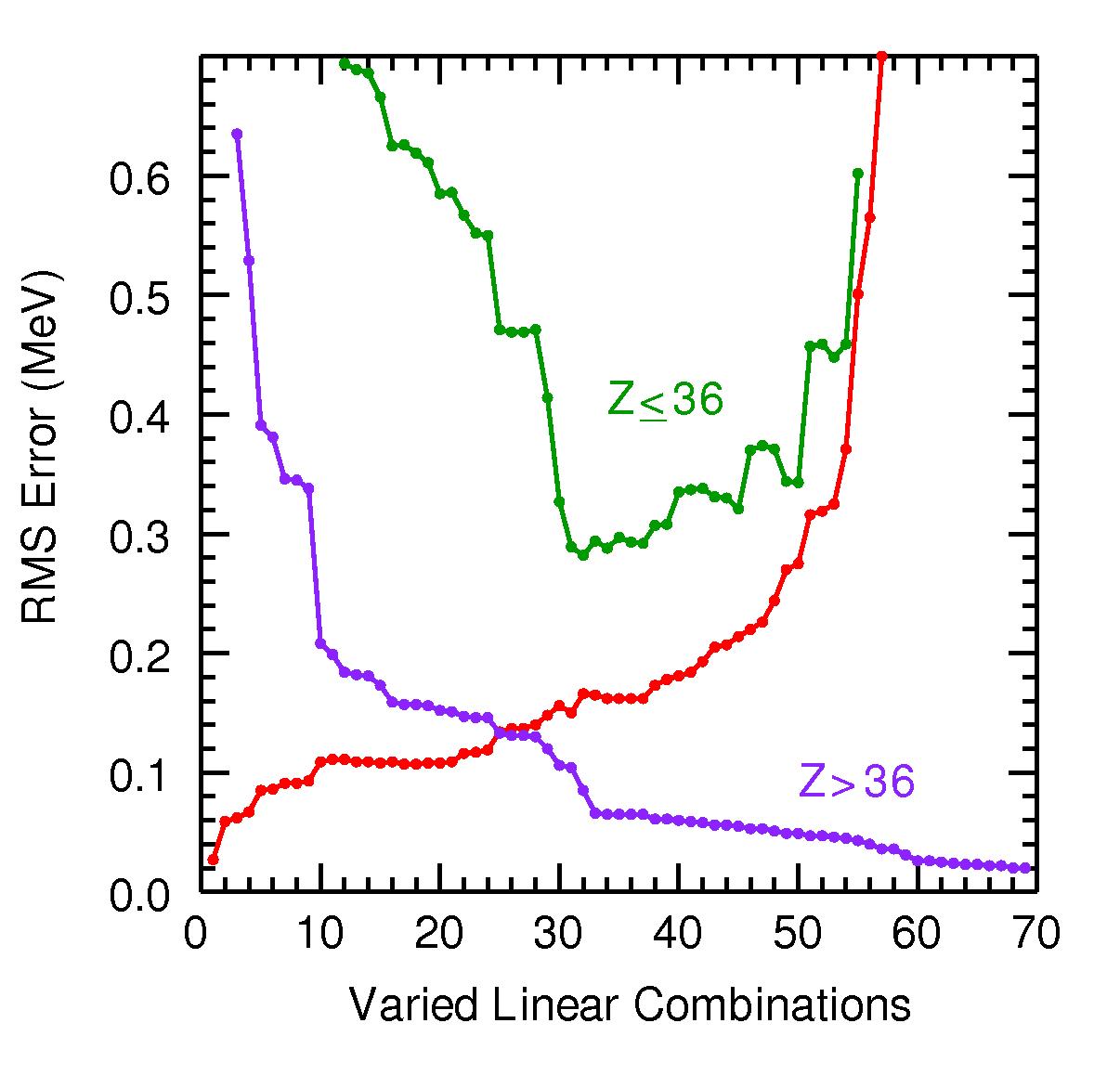}
\caption{Results obtained when the fitted data set is restricted
to $A>86$. See caption to FIG.~\ref{batches}  
}
\label{a86}
\end{figure}

The calculations presented in FIG.~\ref{a86} sampled from the full range of data $28\leq Z \leq 50$, so the predictions are in some sense an interpolation. It is of great interest to also understand the robustness of extrapolations beyond the fit data.
To do this, we partition the data so that the training data only contains $Z>36$, and consider the RMSD for $Z\leq 36$.
We find that the RMSD in the neutron-rich validation set decreases as the number of VLC is increased, up to 35, where we obtain an RMSD of approximately 300~keV.
Beyond 35 VLC, the RMSD grows again, indicating that we have begun to over-fit, and extrapolative power deteriorates.

We found that two markedly different starting Hamiltonians can be tuned to produce quite similar results when a robust dataset and fitting method are implemented. 
The degree to which these starting Hamiltonians must be modified intuitively depends on how
successful the initial parameters are at reproducing the known experimental data. 
The goal is to produce wavefunctions that are realistic and predictive. 
The spectra predicted from our p35-i3 Hamiltonian are compared with their experimental counterparts, 
and in comparison to all experimental data are shown in the Section VII (Appendix). 
%The final fit results and the  TBME for the p35-i3 Hamiltonian are given as text in 
%files in the supplementary material. 
\begin{figure}
\includegraphics[scale=0.52]{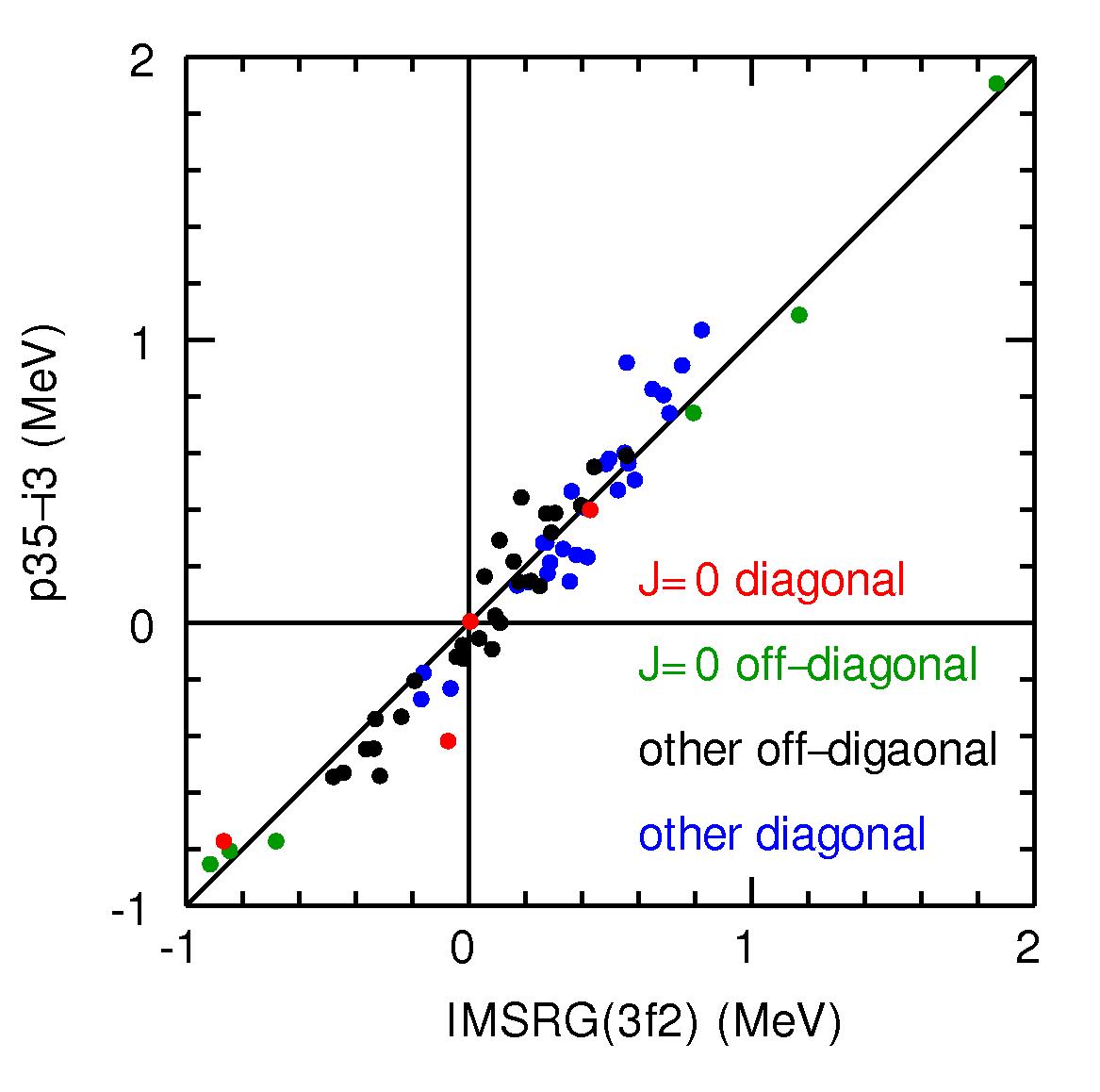}
\caption{Comparison of the IMSRG(3f2) and p35-i3  TBME.
}
\label{initialFinalt}
\end{figure}

%Also figures are given for each nucleus that compare
%all experimental level data with all calculated levels up to six MeV in excitation energy.

\section{Conclusions}
We find that the factorized approximation of IMSRG(3f2) given by \cite{He24} 
produces a starting Hamiltonian that requires less modification to reproduce experimental 
data than ${\it ab initio}$ Hamiltonians produced by other methods, meaning that this Hamiltonian 
is inherently more realistic - a result illustrated clearly in FIG. \ref{batches} and FIG. \ref{initialFinalt}. 
The TBME-RMSD is much smaller with the recently
introduced VS-IMSRG(3f2) corrections involving three-body operators.
Thus, 3f2 should provide a better input for
the linear combinations of parameters that cannot be
determined from VLC fits to energy data.
In addition, we have demonstrated that utilizing the RMSD from the training/validation 
partitioning in the fit procedure helps
protect against over-fitting, and yields more robust extrapolation to data 
beyond that used in the fit. 
The resulting Hamiltonian 
predicts a subset of the experimental spectra for all nuclei in the $\pi j4$ space to 
within a 100-130 keV RMSD. 

The figures in Section VII (Appendix) show the calculated spectra for all
nuclei in comparison to experimnental data.
Many of the predicted theoretical states are not yet observed in
experiment. 
We predict the  
binding energy of $^{100}$Sn \cite{supp}, a doubly-magic nucleus that is of
 imminent experimental interest because it is so proton rich, 
 and lies at the edges of stability. Also one needs
 to confirm our extrapolations for the excited states of $^{99}$In by
 their gamma dcay following one-proton knockout reactions on $^{100}$Sn.
 In the other extreme in $^{79}$Cu, one needs to confirm our extrapolated
 excitation energy for the $9/2^+$ excited state.

We acknowledge support from NSF grants PHY-2110365 and PHY-2340834.

\clearpage

\section*{APPENDIX: COMPARISON OF EXPERIMENTAL AND THEORETICAL ENERGY LEVELS}

%\subsection{Energy Spectra}
Comparison of theory and experiment for all nuclei in the $\pi j4$ model space. 
The color of each line gives the parity of the state (red for odd, 
blue for even) and the length of the line gives the $J$-value of that state. 
States with only a black dot have an experimentally well-determined energy, 
but no definitely assigned $J^\pi$-value. States which have a colored bar 
but with only a black dot at the end have well-determined energy 
and parity, but have a tentatively assigned $J$-value. 
The large circles on the experimental 
levels on the left-hand side are those that were used for the SVD fits with results
shown by the levels with large circles on the right-hand side. 
The black 'X' marks in $^{84}$Se are $0^+$ intruder states coming from 
neutron two-particle two-hole excitations across the $N=50$ neutron shell gap \cite{se84mullin}.

\setcounter{figure}{0}

\renewcommand{\thefigure}{A\arabic{figure}}

\begin{figure}
\includegraphics[scale=0.5]{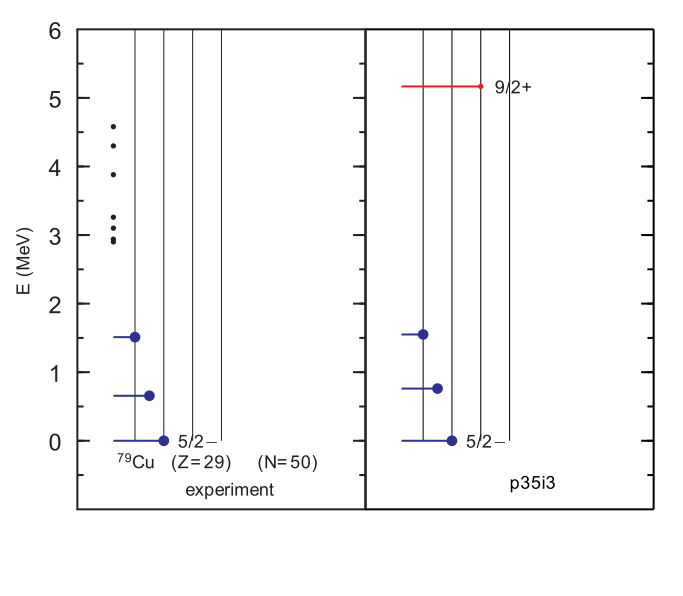}
\caption{Results for $^{79}$Cu.}
\end{figure}

\begin{figure}
\includegraphics[scale=0.5]{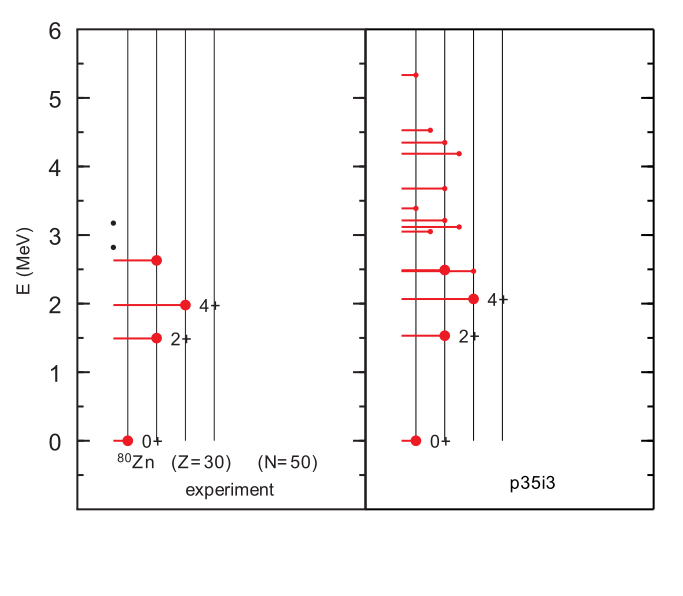}
\caption{Results for $^{80}$Zn.}
\end{figure}

\begin{figure}
\includegraphics[scale=0.5]{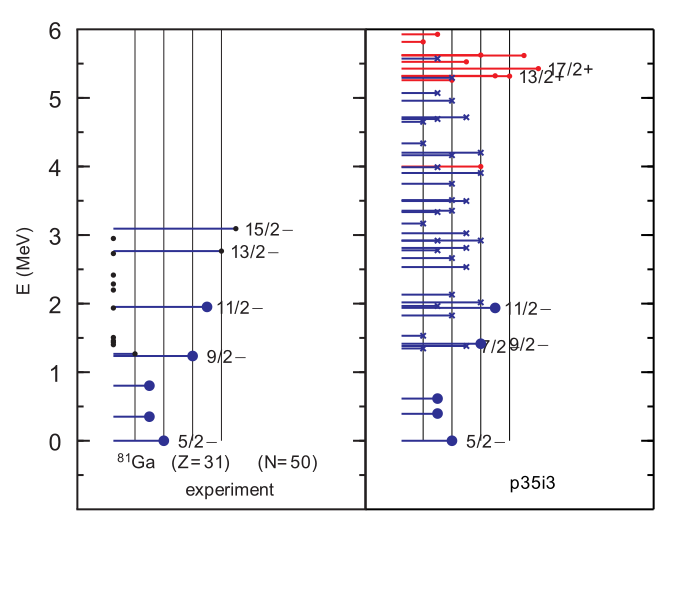}
\caption{Results for $^{81}$Ga.}
\end{figure}

\begin{figure}
\includegraphics[scale=0.5]{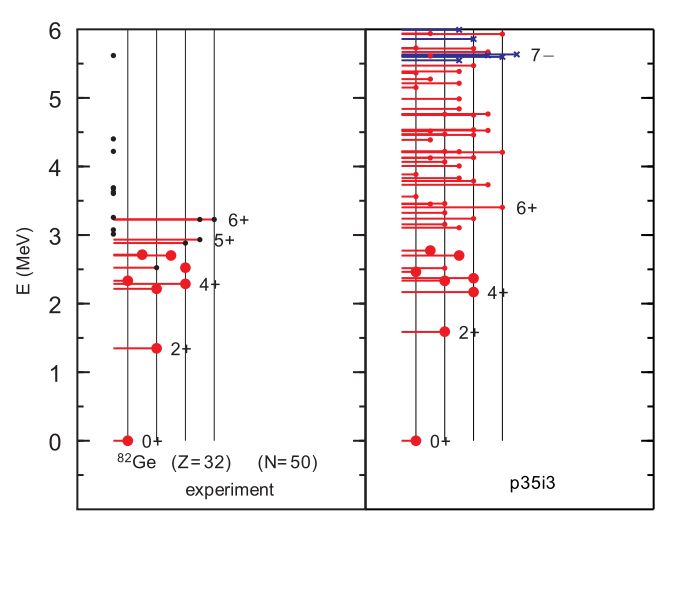}
\caption{Results for $^{82}$Ge.}
\end{figure}

\begin{figure}
\includegraphics[scale=0.5]{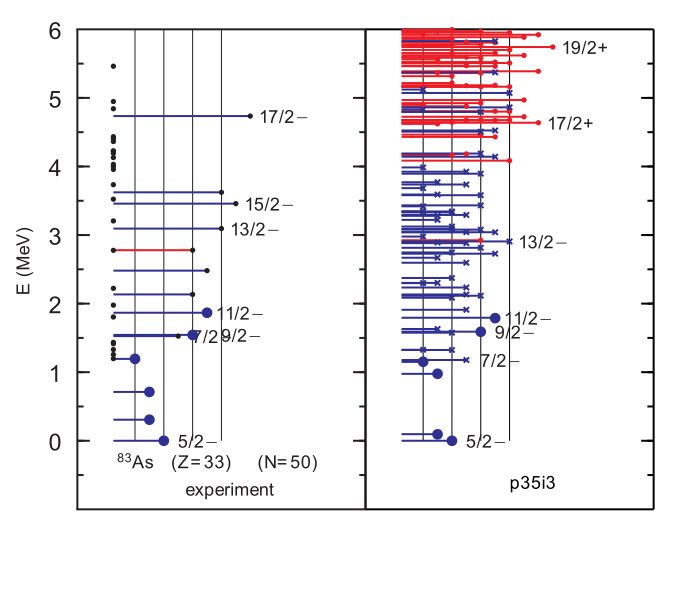}
\caption{Results for $^{83}$As.}
\end{figure}

\begin{figure}
\includegraphics[scale=0.5]{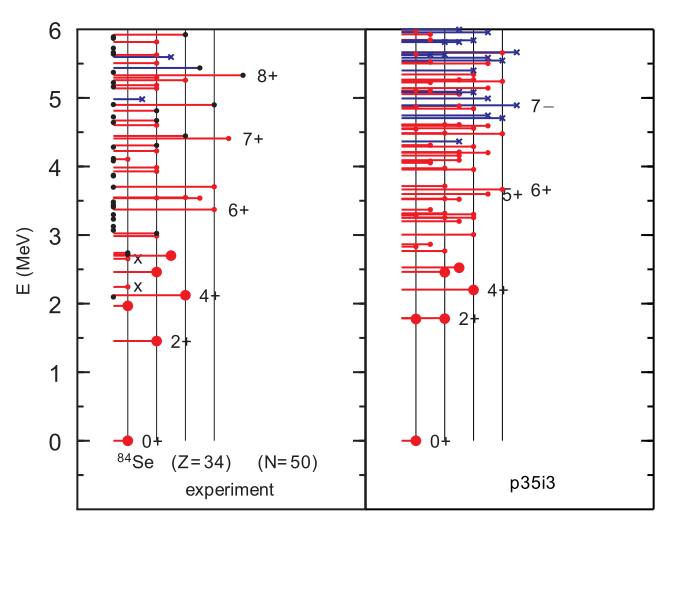}
\caption{Results for $^{84}$Se.}
\end{figure}

\begin{figure}
\includegraphics[scale=0.5]{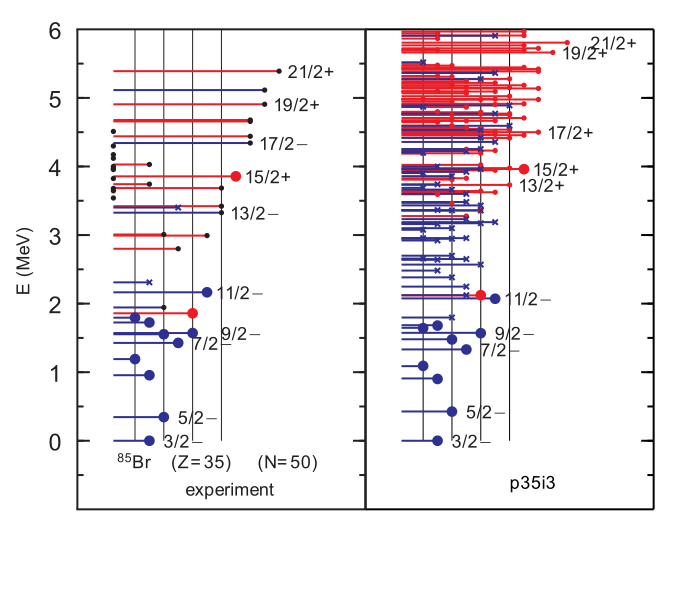}
\caption{Results for $^{85}$Br.}
\end{figure}

\begin{figure}
\includegraphics[scale=0.5]{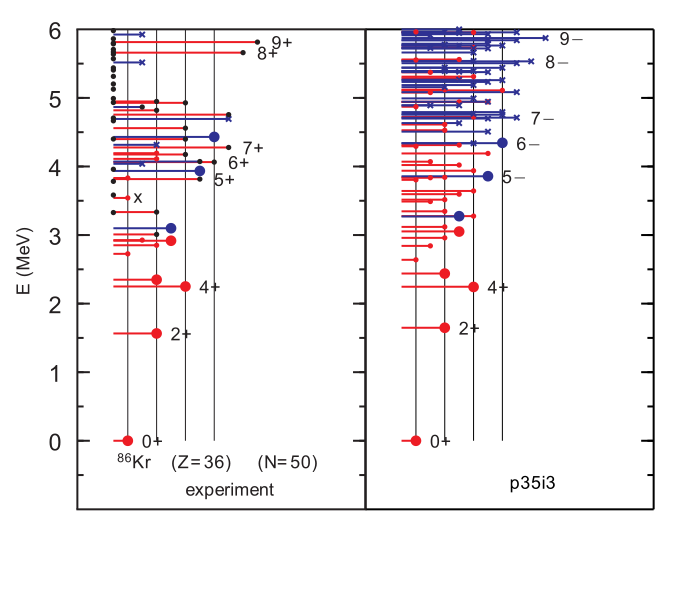}
\caption{Results for $^{86}$Kr.}
\end{figure}

\begin{figure}
\includegraphics[scale=0.5]{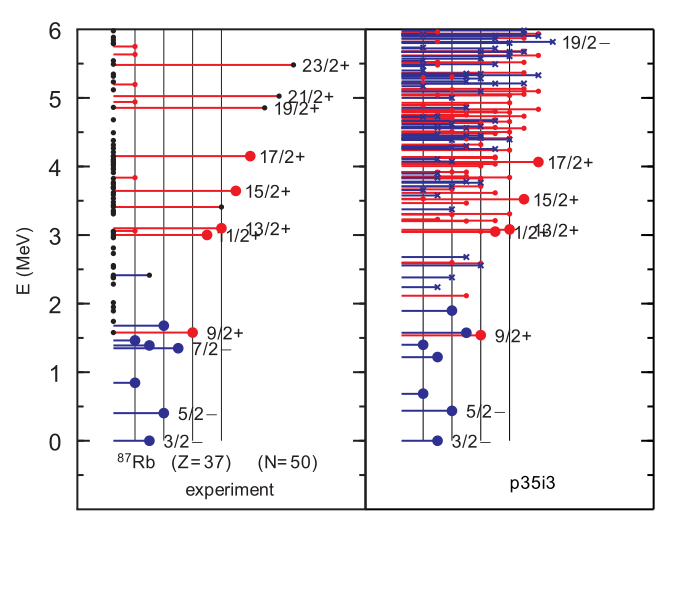}
\caption{Results for $^{87}$Rb.}
\end{figure}

\begin{figure}
\includegraphics[scale=0.5]{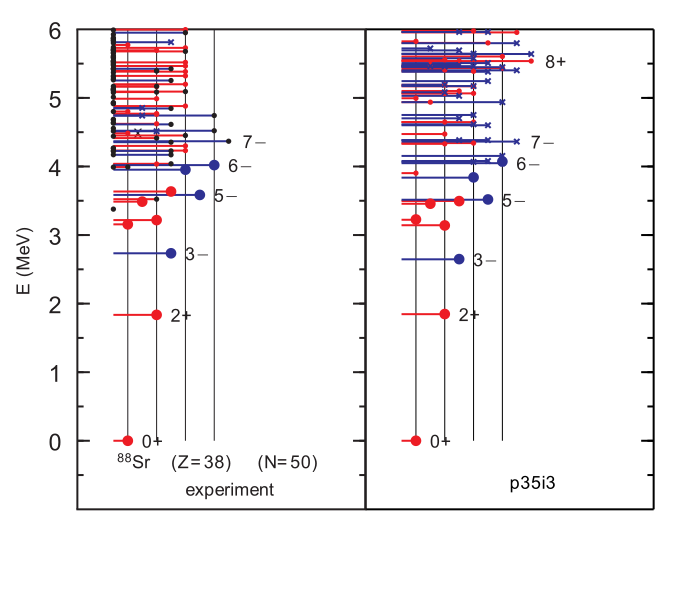}
\caption{Results for $^{88}$Sr.}
\end{figure}

\begin{figure}
\includegraphics[scale=0.5]{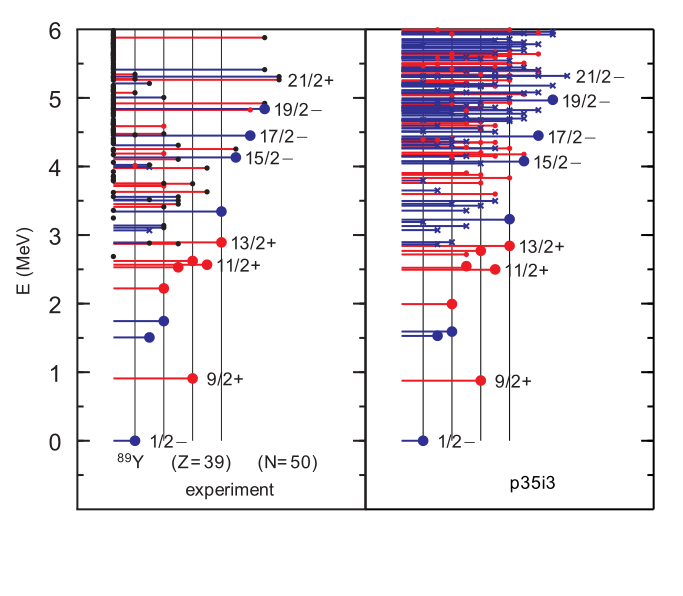}
\caption{Results for $^{89}$Y.}
\end{figure}

\begin{figure}
\includegraphics[scale=0.5]{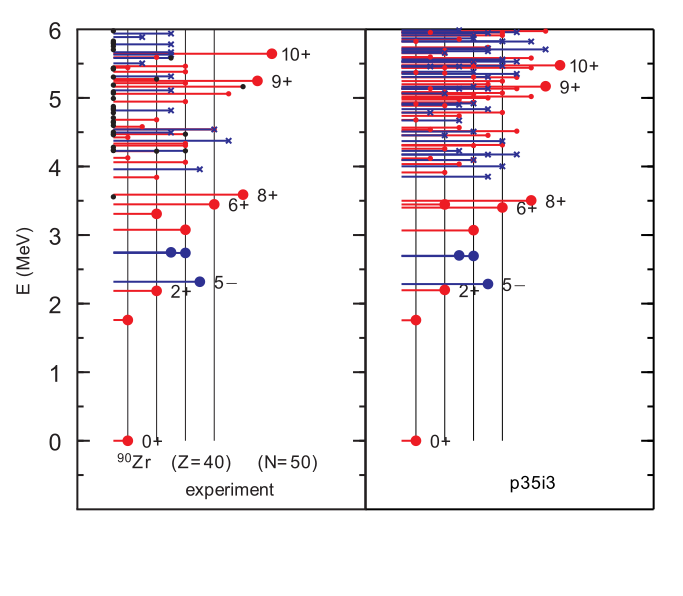}
\caption{Results for $^{90}$Zr.}
\end{figure}

\begin{figure}
\includegraphics[scale=0.5]{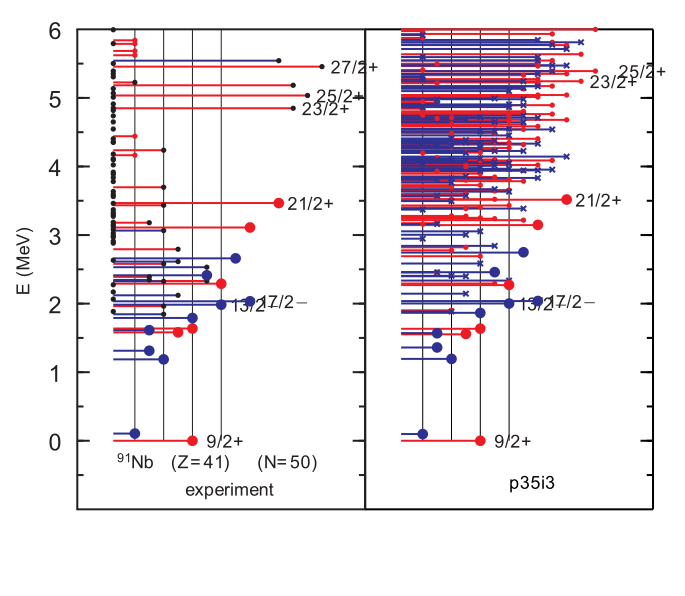}
\caption{Results for $^{91}$Nb.}
\end{figure}

\begin{figure}
\includegraphics[scale=0.5]{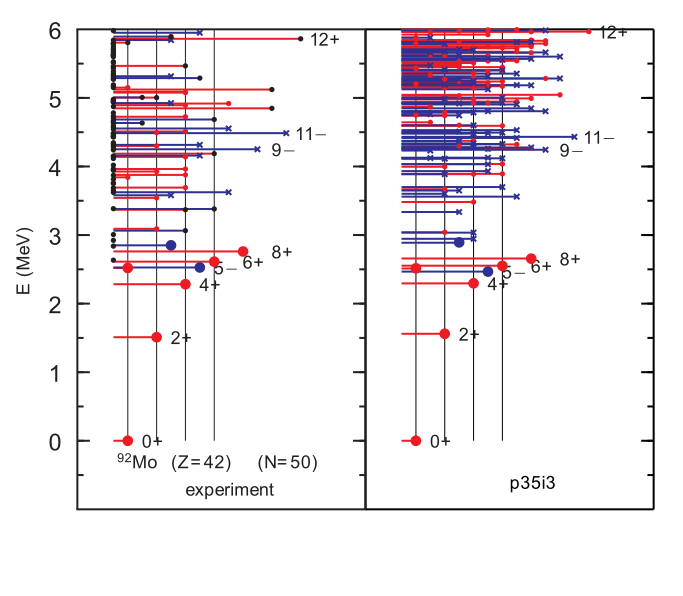}
\caption{Results for $^{92}$Mo.}
\end{figure}

\begin{figure}
\includegraphics[scale=0.5]{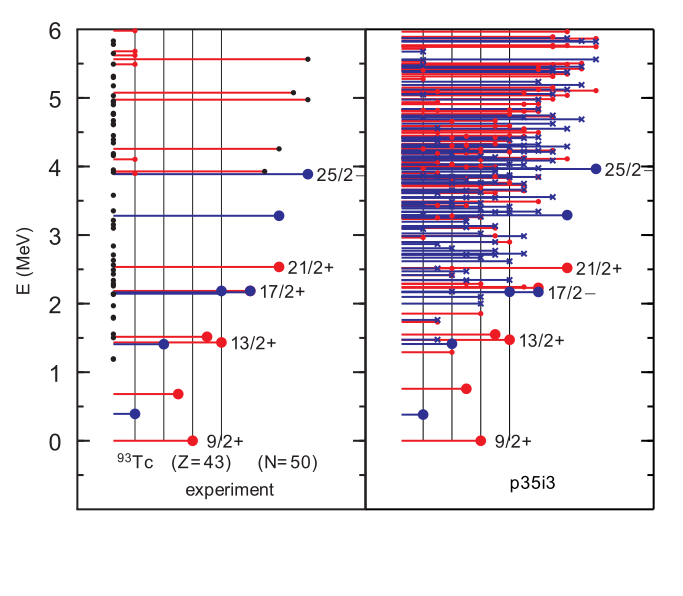}
\caption{Results for $^{93}$Tc.}
\end{figure}

\begin{figure}
\includegraphics[scale=0.5]{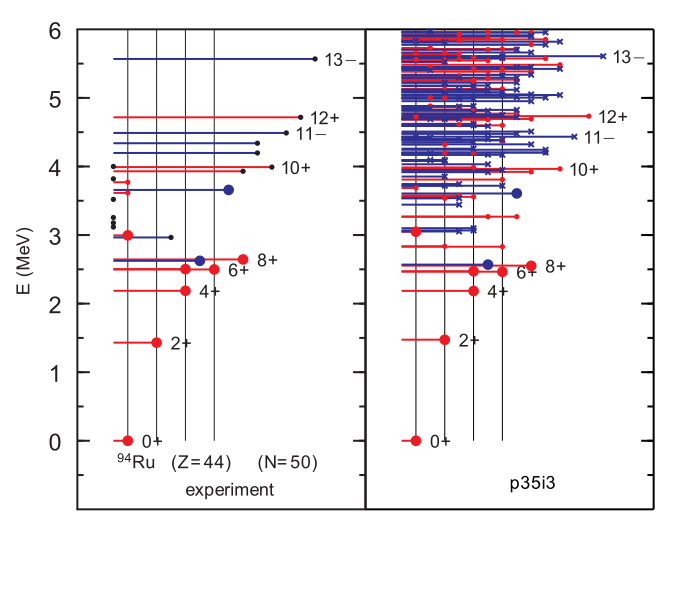}
\caption{Results for $^{94}$Ru.}
\end{figure}

\begin{figure}
\includegraphics[scale=0.5]{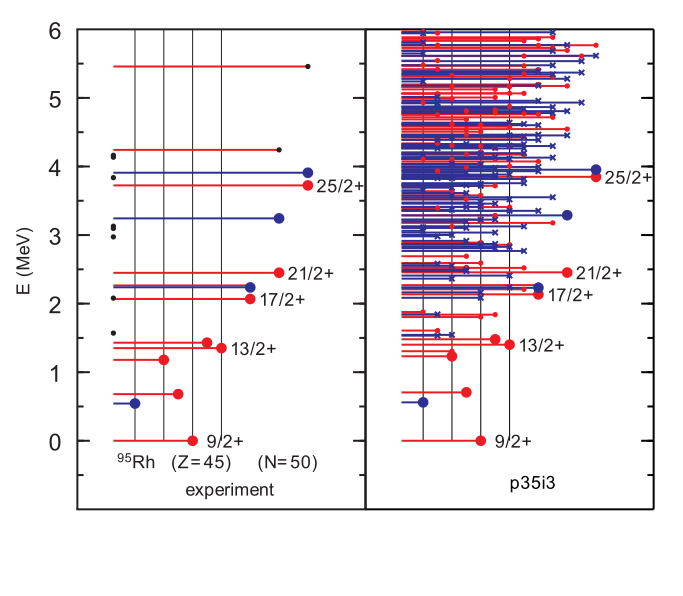}
\caption{Results for $^{95}$Rh.}
\end{figure}

\begin{figure}
\includegraphics[scale=0.5]{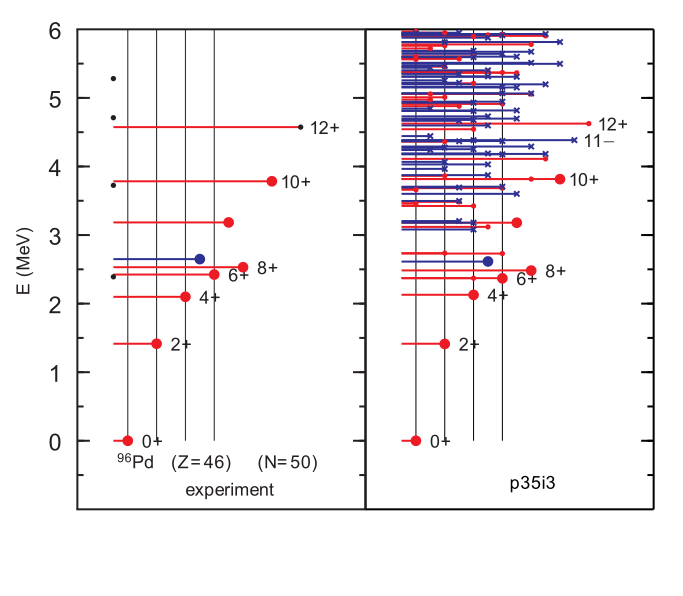}
\caption{Results for $^{96}$Pd.}
\end{figure}

\begin{figure}
\includegraphics[scale=0.5]{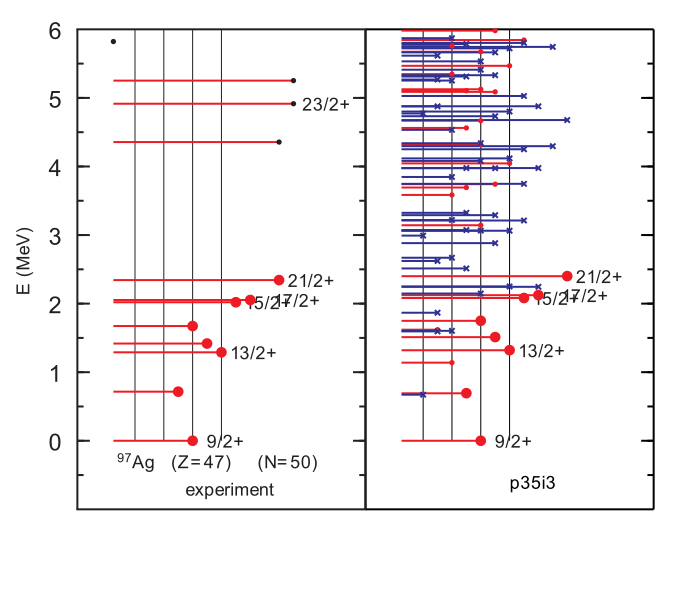}
\caption{Results for $^{97}$Ag.}
\end{figure}

\begin{figure}
\includegraphics[scale=0.5]{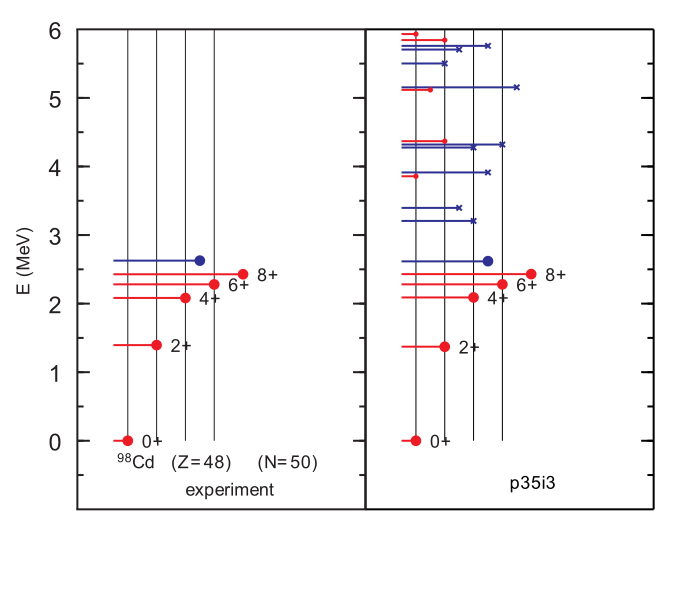}
\caption{Results for $^{98}$Cd.}
\end{figure}

\begin{figure}
\includegraphics[scale=0.5]{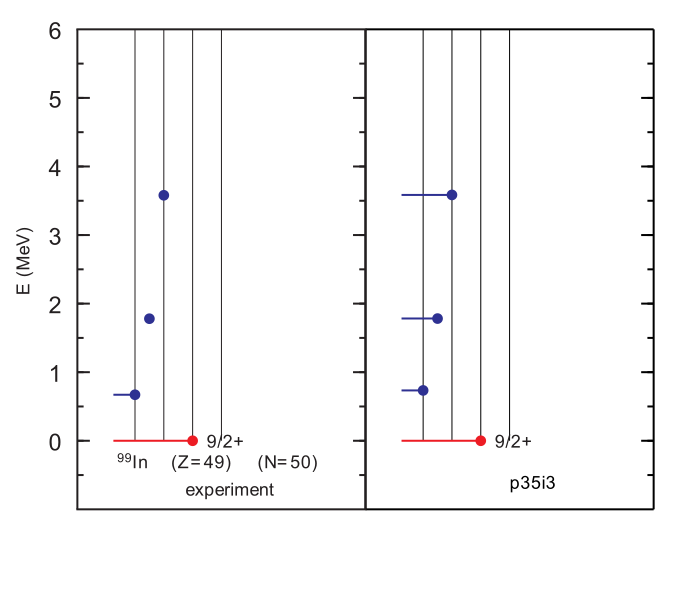}
\caption{Results for $^{99}$In.}
\end{figure}

\end{document}